**New concept for quantification of similarity relates entropy and energy of objects: First and Second Law entangled, equivalence of temperature and time proposed.**

Petr Zimak, Silvia Terenzi and Peter Strazewski*

Université de Lyon, F – 69622, Lyon, France;

Université Claude Bernard Lyon 1, Villeurbanne;

UMR 5246, CNRS, Laboratoire de Synthèse de Biomolécules

*Abstract.* – When the *difference* between changes in energy and entropy at a given temperature is correlated with the *ratio* between the same changes in energy and entropy at zero average free energy of an ensemble of similar but distinct molecule-sized objects, a highly significant linear dependence results from which a relationship between energy and entropy is derived and the degree of similarity between the distinctly different members within the group of objects can be quantified. This fundamental energy-entropy relationship is likely to be of general interest in physics, most notably in particle physics and cosmology. We predict a consistent and testable way of classifying mini black holes, to be generated in future Large Hadron Collider experiments, by their gravitational energy and area entropy. For any isolated universe we propose absolute temperature and absolute time to be equivalent, much in the same way as energy and entropy are for an isolated ensemble of similar objects. According to this principle, the cosmological constant is the squared product of absolute time and absolute temperature. The symmetry break into a time component and a temperature component of the universe takes place when the first irreversible movement occurs owing to a growing accessed total volume.

*Introduction.* – The larger the physical scale is, the less frequently the term 'energy' and the more frequently the term 'entropy' is used in physics discussions. Energy, in the sense of 'bound' or 'inner' energy, is an entity that is usually measured experimentally in some more or less direct way. Entropy is an entity impossible to measure directly; it can only be determined either in conjunction with measured energy and some other measured experimental parameter, free energy for instance, or it is calculated or counted using statistical mechanics or some other theory on the degeneracy of microstates. Since, owing to their distance from the observer, very large-scale physical objects are difficult to measure directly, the preferential use of entropy and the Second Law of thermodynamics is not astonishing in cosmology, neither is the preferential use of energy in quantum physics, in particular, strict energy conservation as expressed through the First Law of thermodynamics. Of course both laws apply a priori to all scales and physics, and of course the above statements are not based on statistical analyses or other objective grounds but on the subjective impression of the author to whom correspondence should be addressed.

In this article we very briefly present the results of a comprehensive analysis of published experimental thermodynamic data on the unfolding of many hundreds of proteins and nucleic acids, on molecular associations in host-guest complexes, on the stability of ab initio (quantum mechanically) calculated water clusters and the semi-empirically (force field) calculated formation thermodynamics of small organic molecules from their elements. We then mainly discuss the consequences when i) these numerical results are first grouped into families that distinguish ensembles of evidently similar objects, ii) the grouped results are correlated in a specific two-dimensional projection of a five-dimensional parameter space, iii) then detached from the molecular scale and, ultimately, iv) from the physical identity of the initially used parameters and variables. We finally propose to take a fresh look at possible interpretations in physics of the combination of the mathematical operations of addition and multiplication, in order to attempt to reduce the number of axiomatic assumptions that govern different physical models at very different scales.

The discussion begins with deriving an equation that relates for the first time energy changes to entropy changes of the same objects without usage of additional empirical parameters or functions that are not explained from the fundamentals. The only new 'entity' or 'information' is the fact that the objects are grouped into families of obviously similar characteristics. Protein mutants and nucleic acid variants are macromolecules that usually differ only very little in overall shape and folding potential – only one or two in dozens or hundreds of 'chain links' are different within the same group – but may differ rather heavily in measured energy and entropy of folding. It is known since 1970 that in many very different chemical and biological systems large entropy and energy contributions compensate each other, to give small resulting free energy changes, that is, small net effects. We do not discuss this here – our studies on the compensation effect are described in full detail to be published elsewhere – but rather focus on the fact that, once energy and entropy changes are fundamentally linked to one another, the laws that on the one hand restrict in isolated systems average net energy changes to zero and on the other hand confine spontaneous net entropy changes to zero or more but not less, thus, condemn entropy to maximize over time, may become fundamentally linked as well. If our analysis on the thermodynamics of medium-sized objects, which can either be described by quantum physics or by classical physics, were generalizable to all scales, we were to conclude the following.

The First and Second Law of thermodynamics describe isolated multicomponent systems in the observable universe as objects that conserve their energy due to their very isolation *and* that spontaneously maximize their entropy over time when their size is such that fully reversible changes within these objects, that is, exactly reversed changes in their microstates, are too improbable to occur within the actual lifetime of the universe. Additionally, an *isolated ensemble of similar objects* in the same universe will spontaneously maximize its overall entropy over time *in a way (at a rate)* that reflects its overall energy *and* identity, i.e., its compositional and structural characteristics that define it as an ensemble of similar objects. If the physical isolation of the ensemble confines its overall average energy changes to zero, the way (rate) of maximizing entropy can only change when the degree of similarity within the ensemble of objects changes as well. To the best of our knowledge, black holes differ from one another through the least of characterizing parameters, mass and angular momentum (and for some time charge) only, whereas elementary particles differ from each other through a whole plethora of characteristics (according to the standard model) and the variability, thus, potential dissimilarity of objects that are composed of these elementary particles multiplies, i.e., increases at a geometric rate with the number of involved particles. We conclude that, given a constant (accessed) overall volume of an ensemble, the higher the degree of similarity is among its objects the slower is their rate of spontaneous entropy maximization and the closer to maximum entropy they are. Hence, it seems as if the rate of maximizing overall entropy of an ensemble of objects were related to the similarity of what characterizes the individual objects within the ensemble.

Here we present a statistical means of quantifying the degree of similarity, namely, through the linear regression coefficient obtained from the correlation of the difference with the ratio of two object characterizing parameters (energy $U$ and entropy $S$) that both depend on one independent variable (absolute temperature $T$). We depict, using experimental numerical values, 3D projections of the 5D parameter space $\{U;\ S;\ T;\ U - T \cdot S;\ U/S\}_{pV}$ (at constant pressure and volume $pV$). We attempt to start a productive discussion in suggesting to treat the whole visible universe as one isolated ensemble of 'similar objects' and to analyze it by using the mathematically same or akin parameter space but replacing the dimensions by, respectively, {absolute temperature $T$ ; absolute time $t$ ; total volume $V$ ; $T + V \cdot t$ ; $T \cdot t$ } where $V$, instead of $T$, is the 'independent variable'.

*Methods*. – The vast majority of the primary data are experimental and about one third of those originate from differential scanning calorimetric experiments where, both, the energy change under constant pressure, i.e., the enthalpy change $\Delta H = H_{\text{macrostate 1}} - H_{\text{macrostate 2}}$ in an open, with respect to atmospheric pressure, system (energy $U = H - p \cdot V$), and the position of thermodynamic equilibrium between two macroscopic states, i.e., free enthalpy (Gibbs free energy) change $\Delta G$ as a measure for the driving force towards macroscopic stasis under constant pressure (free energy $F = G - p \cdot V$), are derived from the measured heat capacity $C_p$ (at constant pressure) as a function of $T$ within a $T$-range needed to observe both major macroscopic states (termed 'folded' and 'unfolded') in virtually quantitative abundance, and using equation 1. The corresponding change in entropy $\Delta S$ is usually calculated from $\Delta G = \Delta H - T \cdot \Delta S$, rather than directly from equation 1.

$$C_p = dH/dT = T \cdot dS/dT = -T \cdot (d^2G/dT^2) \qquad (1)$$

Another definition of heat capacity is the mean squared fluctuation in energy scaled by $kT^2$, or the mean squared fluctuation in entropy scaled by $k$ (the Boltzmann constant), as shown in equation 2 [1].

$$C_p = \langle \delta H^2 \rangle / kT^2 = \langle \delta S^2 \rangle / k \qquad (2)$$

The difference in specific heat capacity between both major macroscopic states is directly measured from $\Delta C_p = C_p(T_{100\% \text{ unfolded}}) - C_p(T_{100\% \text{ folded}})$ ($\Delta$ always refers to the difference between two distinct macroscopic states) where both $C_p$(100% unfolded) and $C_p$(100% folded) are assumed to exert the same $T$-dependence, hence $\partial \Delta C_p / \partial T = 0$, i.e. $\Delta C_p \approx \text{const}$. The other two thirds of experimental data originate from so-called van't Hoff experiments in which, instead of $C_p$, equilibrium constant $K = (\text{fraction macrostate 1})/(\text{fraction macrostate 2}) = \exp[-\Delta G/RT]$ ($R = 1.9872$ cal mol$^{-1}$ K$^{-1}$) is measured within an appropriate range of $T$ or other parameter capable of completely shifting the thermodynamic equilibrium from one macroscopic state to another. For thermally induced macrostate changes the accompanying energy and entropy changes are elucidated from fitting the experimental data to equation 3:

$$R \cdot \ln K = -\Delta H / T + \Delta S = -\Delta G / T \qquad (3)$$

In the vast majority of published van't Hoff experiments heat capacity changes are ignored altogether: $\Delta C_p \approx 0$. This approximation is justified by the usually observed linear relationship for $\ln K$ versus $1/T$. In both kinds of experiments, calorimetric and van't Hoff, any true $T$-dependence of $\Delta C_p$ may be neglected when compared to the one of $\Delta G = \Delta H - T \cdot \Delta S$ (or of $\Delta F = \Delta U - T \cdot \Delta S$) over the measured $T$-range. In summary, classical thermodynamics provides us with equations 4 and 5 in the fundamental, most general case $\Delta C_p = f(T)$ [2]. Equations 6 and 7 result from the 'calorimetric neglection' of the $T$-dependence of $\Delta C_p$. After a 'van't Hoff neglection' of $\Delta C_p$, $\Delta H$ and $\Delta S$ become constants with respect to $T$.

$$\Delta H_T = \Delta H_{T_{ref}} + \int_{T_{ref}}^{T} \Delta C_p(T) dT \quad ; \quad \Delta S_T = \Delta S_{T_{ref}} + \int_{T_{ref}}^{T} \frac{\Delta C_p(T)}{T} dT \qquad (4) ; (5)$$

$$\Delta H_T = \Delta H_{T_{ref}} + \Delta C_p \cdot (T - T_{ref}) \quad ; \quad \Delta S_T = \Delta S_{T_{ref}} + \Delta C_p \cdot \ln(T/T_{ref}) \qquad (6) ; (7)$$

***Procedure***. – We extracted from the literature [3] 1555 experimental datasets {$\Delta C_p$ ; $\Delta H_{T_{ref}}$ ; $\Delta S_{T_{ref}}$} on the thermal and non-thermal unfolding of proteins and nucleic acids, where for each dataset $T_{ref} = T_{\Delta H = T \cdot \Delta S} = T_m$. $T_m$ is the so-called midpoint or equilibrium temperature, the temperature at which in a dynamic and fully reversible two-state equilibrium the fractions of both (two particularly stable and well observable) macrostates are equal, therefore $\Delta G_{T_m} = 0$ (eqn. 3). We expanded the above datasets with an additional function each, the state function $\Delta G_T = \Delta H_T - T \cdot \Delta S_T$, using equations 3 (right-hand side), 6 and 7. At that stage, no numerical values were attributed to $T$ yet. Each dataset was now made up of five 'characterizing parameters' {$\Delta C_p$ ; $\Delta H_{T_m}$ ; $\Delta S_{T_m}$ ; $T_m = \Delta H_{T_m}/\Delta S_{T_m}$ ; $\Delta G_T = \Delta H_T - T \cdot \Delta S_T$}, *all of which are dependent on one another through the fundamental thermodynamic equations 1 to 5*, and of one 'independent variable' $T$. Note that all five parameters, despite being derived from $C_p$ and $T$, bear distinct physical meanings (interpretations).

All 1555 datasets were then grouped into 154 families, according to the structural similarity of the members within each group (mostly 'single-chain link' variants, 'point mutants'). The datasets of each of the 154 groups were submitted to a group-specific correlation between the two combined (with respect to $\Delta H$ and $\Delta S$) parameters $\Delta G_T$ and $T_m$. An increasingly refined sampling of $\Delta G_T$ on a representative part of the groups led to a complete correlation analysis $\Delta G_{T_{median}}$ vs. $T_m$ of all groups at a group-specific $T = T_{median}$. $T_{median}$ is the statistical median of all equilibrium temperatures $T_m$ of a group.

***Results***. – The correlations between $T = 273$ and $373$ K appeared visibly linear for the vast majority of the analyzed groups, hence, a linear regression according to equation 8 was used to characterize every group.

$$\Delta G_T = h_T - T_m \cdot s_T = h_T - (\Delta H_{T_m}/\Delta S_{T_m}) \cdot s_T \qquad (8)$$

For a detailed consultation of the results see the Supporting Information (numerical and graphical SI and Mathematical Appendix). Here it suffices to note that all members of the same group share the same 'group parameters' $h_T$ and $s_T$ which express nothing more than the average energy and, respectively, entropy of the group of similar objects. They are therefore only dependent on $T$ and the choice of which individual members constitute 'a group'. The numerical values for the slope $s_{T_{median}}$ are actually average values of all numerical $\Delta S_{T_m}$ values of each group member within one group. The numerical values for $h_T$ and all other $s_T$ depend on $\Delta C_p(T)$, the more so the larger $|T - T_{median}|$ is. According to equation 8 the $T$-dependence of $h_T$ and $s_T$ is the same as for $\Delta G_T$. For $\Delta C_p =$ const. this $T$-dependence adopts the form $f(T) = a + b \cdot T + c \cdot T \cdot \ln T$, in which c is nil for $\Delta C_p = 0$ (eqns. 3, 6 and 7). We fitted this function to all experimental data, to obtain the 'group constants' (with respect to $T$) $h_{0-2}$ and $s_{0-2}$ for $h_T = h_0 + h_1 \cdot T + h_2 \cdot T \cdot \ln T$ and $s_T = s_0 + s_1 \cdot T + s_2 \cdot T \cdot \ln T$. Note that $h_{0-2}$ and $s_{0-2}$ can all be derived from the $\Delta S_{T_m}$, $\Delta C_p$ and $T_m$ values of a group (SI: equation 42 in the Mathematical Appendix).

The main and new result is that at $T_{median}$, at the temperature where the sum of $\Delta G$ of all group members within one group is closest to nil, the vast majority of experimental data produces a linearity of unexpected quality. The linearity as such remains visible but its quality, as expressed through the regression coefficient, degrades quite strongly and monotonously with increased $|T - T_{median}|$ (SI : Figs. S14-S15) and, in a non-trivial fashion, as we join evidently less similar objects into the analyzed group (Figs. S1, S5-S6, S10-S11). The experimental group sizes vary between 4 and 68 (average 10). The regression coefficients $r_{T_{median}}$ of all calorimetric groups lie between 0.90 and 0.999'999 with an abundance maximum between 0.999 and 0.9999 (Figs. S12-S13). The van't Hoff groups do not fall far behind (Fig. S7). In addition, the same correlation method was tested on the calculated thermodynamics of formation from the pure chemical elements in their standard state of a homologue series of PM3-calculated simple organic molecules, as well as of published ab initio-calculated water clusters [4], using statistical thermodynamics at 298 K. The somewhat lower correlation coefficients $r_{298K}$ as compared to the above experimental $r_{T_{median}}$ values are due to the fact in part that at $T = 298$ K many calculated data points within one group do not center around $\Delta G = 0$. The linearity of similar groups is nevertheless unambiguously apparent (Figs. S37-S39).

***Discussion***. – The mere fact that changes in energy and entropy are fundamentally correlated is not unexpected – after all, their temperature dependence is akin and dictated by the corresponding change in heat capacity (eqn. 1), i.e., their mean fluctuation (eqn. 2) – neither is the relatedness between free energy and the temperature at which it vanishes. Both $\Delta G_T$ and $T_m$ are commonly interpreted as a representation of 'thermodynamic stability', the former is expressed in energy units and depends on $\Delta C_p(T)$, the latter lends its unit from the temperature scale and is untouched by any $T$-dependence of $\Delta C_p$. However, this linearity has never been systematically demonstrated from experimental data, nor its strong dependence on the similarity of congeners, nor its highest quality at $T = T_{median}$ and so far it has never been formulated as a consequence, a necessity imposed by the theory of thermodynamics. The distinct linear grouping of the theoretically calculated molecules (of chemically very different nature from that of proteins or nucleic acids) is at least inasmuch significant as the thermodynamic parameters are, despite the not entirely exact nature of their elucidation (due to the harmonic oscillation approximation for calculating $S$), independently derived from partition functions and not from experimental enthalpies or experimental equilibrium constants.

Taken together, the similarity-dependent linearity of $\Delta G_{T_{median}}$ vs. $T_m$, quantified through the regression coefficient $r_{T_{median}}$, seems to be as general as the whole theory of thermodynamics is. Therefore we proceed with deriving general consequences such as the entanglement of the First and Second Laws for groups of similar objects, as mentioned in the Introduction, or the analysis of a function that was generated from the combination of equations 3, 8 (both right-hand side), 4 and 5 to give through the elimination of $\Delta G_T$ equations 9 and 10, i.e., the fundamental energy-entropy relationship and mathematical basis for the 5D parameter space $\{\Delta H_{T_m} ; \Delta S_{T_m} ; T_m = \Delta H_{T_m}/\Delta S_{T_m} ; \Delta G_T = \Delta H_T - T \cdot \Delta S_T ; T\}$. Equation 9 is a simplified version for $\Delta C_p = 0$ (for clarity) of the general form shown as equation 10. Both equations can be analytically solved for $\Delta S_{T_m}$ (SI: equation 26 in the Mathematical Appendix).

$$\Delta H_{T_m} = T \cdot \Delta S_{T_m} \cdot \frac{\frac{h_T}{T} + \Delta S_{T_m}}{s_T + \Delta S_{T_m}} \tag{9}$$

$$\Delta H_{T_m} = T \cdot \Delta S_{T_m} \cdot \frac{\frac{h_T - \int_{T_m}^{T} \Delta C_p(T) dT + T \cdot \int_{T_m}^{T} \left(\frac{\Delta C_p(T)}{T}\right) dT}{T} + \Delta S_{T_m}}{s_T + \Delta S_{T_m}} \tag{10}$$

The above functions are variants of the well known quadric $z = x \cdot y$ of the shape of a hyperbolic paraboloid, thus, a single saddle point centered in the origin and the $S_4$-symmetric function spreading from there with an all-negative Gaussian curvature $K = -(1 + x^2 + y^2)^{-2}$. For $\Delta C_p = 0$ (eqn. 9 with $h_T = h_0 + h_1 \cdot T$ and $s_T = s_0 + s_1 \cdot T$ from the van't Hoff datasets) the basic shape of the function does not change when compared to $z = x \cdot y$, although the function area may be quite heavily distorted (not shown). However, for $\Delta C_p \neq 0 =$ const. (eqn. 9 with $h_T = h_0 + h_1 \cdot T + h_2 \cdot T \cdot \ln T$ and $s_T = s_0 + s_1 \cdot T + s_2 \cdot T \cdot \ln T$) the group constants $h_{0-2}$ and $s_{0-2}$ that were obtained from the experimental calorimetric datasets produced shapes of the eyebrow-rising kind. In Figure 1 four views of the same 3D-projection, $\Delta H_T$ versus $\Delta S_T$ and $T$, of the thermodynamic 5D parameter space is shown for one particular but representative calorimetrically measured protein mutant group (mutants of Staphylococcal Nuclease). In Figure 2 one to two views of three more 3D-projections for the same mutant group are depicted. Both figures focus on the zone that contains the experimental data.

The interested reader is welcome to copy from the numerical Supporting Information file any set of experimental group constants $h_{0-2}$ and $s_{0-2}$, plot equation 9 at any scale (solved for $\Delta S_{T_m}$ to suppress in certain 3D projections a maximum of asymptotic planes) and enjoy the shapes and wormholes created by the $T \cdot \ln T$ terms. A more comprehensive study on the characteristics of this function shall be published elsewhere. More important for physics is the fact that group specific thermodynamic parameter spaces depict the only possible values that can be realized by a particular group of similar objects. The rest is void, *terra incognita* for the group members, unless an object changes its characteristics (structure, composition, etc.), unless it 'dissimilarizes' off from 'its' group – to join some other one.

*Conclusion*. – Theories from quite different domains such as, to name a few, probability theory [5], information theory and the emergence of complex systems [6], quantum relativity/cosmology [7] and string theory [8] operate with entropy and the Second Law of thermodynamics yet in conjunction with parameters different from the ones studied here. Urgent problems are being at least attacked, and possibly solved, through the insight into apparent and/or fundamental analogies between statistical thermodynamics and, for example (respectively), randomness of sequential irregularities ("algorithmic entropy", "approximate entropy"), computational compactness ("logical depth"), quality change of hereditary information (change in systemic "knowledge" through periodically discarded "Shannon entropy"), the dynamics of black holes ("Bekenstein-Hawking entropy") and tracing back the microscopic origin of their area-entropy by counting – provided the black holes are highly charged – the degeneracy of their periodical and persistent topological defects (Bogomol'nyi-Prasad-Sommerfield soliton bound states). In all above cases the problem arises of how to reliably quantify or sample randomness, logical depth, knowledge, entropy, in order to understand their physical origins and their development over time. Whereas in thermodynamics the absolute temperature $T$ is the (mathematically) 'independent variable' that accompanies the maximization of entropy, it may be time $t$ in Darwinian processes of knowledge maximization or surface gravity $\kappa$ during a life time of an extremely dense cosmic object.

Common to all these theories is the utilization of Legendre(-Fenchel) transformations, for example, to describe various thermodynamic potentials in classical thermodynamics, to derive partition functions from a state equation in statistical thermodynamics, to link Lagrange (classical) mechanics with Hamiltonian (quantum) mechanics, to formulate quantum electrodynamics and quantum chromodynamics, and more. In a *thermodynamic context* the linking of energy with entropy through a quantified similarity argument may be viewed as a Legendre transformation on a group, rather than individual, level. One new observation here is that, for a sufficiently large number of objects, the absolute *linear regression coefficient* obtained from the correlation between (any) two object characterizing entities that are derived from the sum (difference) and product (ratio) of more fundamental entities can be used to *quantify the similarity* between these objects under conditions (choosing the 'independent variable') such that the sum of sums (differences) is closest to nil. We expect this correlation method to be testable in experimental systems at very different scales from the molecular one, be it in size or energy or both. For example, if mini black holes could be transiently generated in future Large Hadron Collider experiments and different classes of such objects could be observed, we would predict that the relationship between their gravitational energy and the surface area of their event horizon would correlate in a fashion that were characteristic for their kind: Energy and entropy would correlate, through equation 10, differently, i.e., with different group parameters for objects of a particular (range of) angular momentum and charge than for another. Distinct groups should appear and be best visible in free energy correlations as formulated in equation 8.

    More generally, we are wondering whether the formal procedure utilized here on thermodynamic parameters, i.e., the analysis of parameter spaces that include dimensions in which the arithmetic operations *addition* and *multiplication* have been separately applied to these parameters, could not be useful elsewhere. *A priori* there is no reason to assume that equations 8, 9 and 10 should be restricted to energy and entropy, since the equations essentially describe the relation between any two object-characterizing 'entities' independent of their physical meaning, as long as these entities can be physically meaningfully related through both the sum and the ratio between them. We may then postulate through a mathematical analogy that the similarity within an ensemble of similar objects will always uncover and be visible in a linear relation between the sum and the ratio of these very entities, irrespective of whether they describe a thermodynamic state or some other state. For instance, a most characteristic feature of quantum physics (including all versions of string theory) is the superposition (in time) of multi-dimensional wave functions, that are basically weighed *sums* of what is interpreted as existence probability densities (in space), which by their very nature are *multiplied* in a series of events over time. Or, what is viewed as an 'independent variable' in thermodynamics, absolute temperature $T$, finds its *alter ego* in general relativity: the total volume $V$ of the universe. Oddly, both 'independent' variables depend during the evolution of the universe on all other parameters in space-time (e.g. energy density, mass density)

and on one another. What seems to happen in an expanding universe due to its increase in total volume $V$ is the spontaneous transformation of inherently reversible particle movements, as expressed through high absolute temperature and early absolute time, into increasingly irreversible ('directed') particle movements, as expressed through lower absolute temperature and later absolute time. Hence, similarly to Einstein's mass-energy equivalence, we postulate equation 11 from the cosmologically observed equivalence of absolute time $t$ and absolute temperature $T$ – lending the units for the latter from the equation of state (squared characteristic thermal speed of particles over universal gas constant) – where $\Lambda$ is the cosmological constant:

$$t \cdot T = \Lambda^{-1/2} \qquad (11)$$

$$\lim_{V \to 0} t = T^{-1} \qquad (12)$$

Within an extremely small (close to the Planck) total volume all movements of all particles are exactly reversible, there is no way at all for any change to manifest itself at any given (extremely small) $V$, and so time and temperature are exactly equivalent (eqn. 12). The symmetry break – most likely the first of all – happens when $V$ becomes such that the first precise back movement is too improbable to occur. At such a critical $V$, time becomes first manifest through not reversed particle movements and thus becomes distinguishable from temperature. We now ask ourselves: If $\Lambda = (t \cdot T)^2$, what is $T + V \cdot t$ ?

*Acknowledgements*. – We thank Prof. Peter Schuster, Theoretische Chemie, Universität Wien, Prof. Emmerich Wilhelm, Physikalische Chemie, Universität Wien, and Prof. Irene Poli, Statistical Department, University Cà Foscari, Venezia, for critically reading an extended version of the manuscript, and Prof. Günter von Kiedrowski, Bioorganische Chemie, Ruhr-Universität Bochum, for critically reading many versions of the manuscript and important enlightening discussions about a Unified Law of Thermodynamics. We are indepted to Prof. Bertrand "BOP" Castro (ex Sanofi-Aventis, Gentilly), for calculating the formation thermodynamics of simple organic homologues, and to Prof. Hans-Christoph Im Hof, Mathematical Institute, University of Basel, for performing a differential geometry analysis on the Gaussian curvature and geodesics of $z = x \cdot y$.

* E-mail: strazewski@univ-lyon1.fr

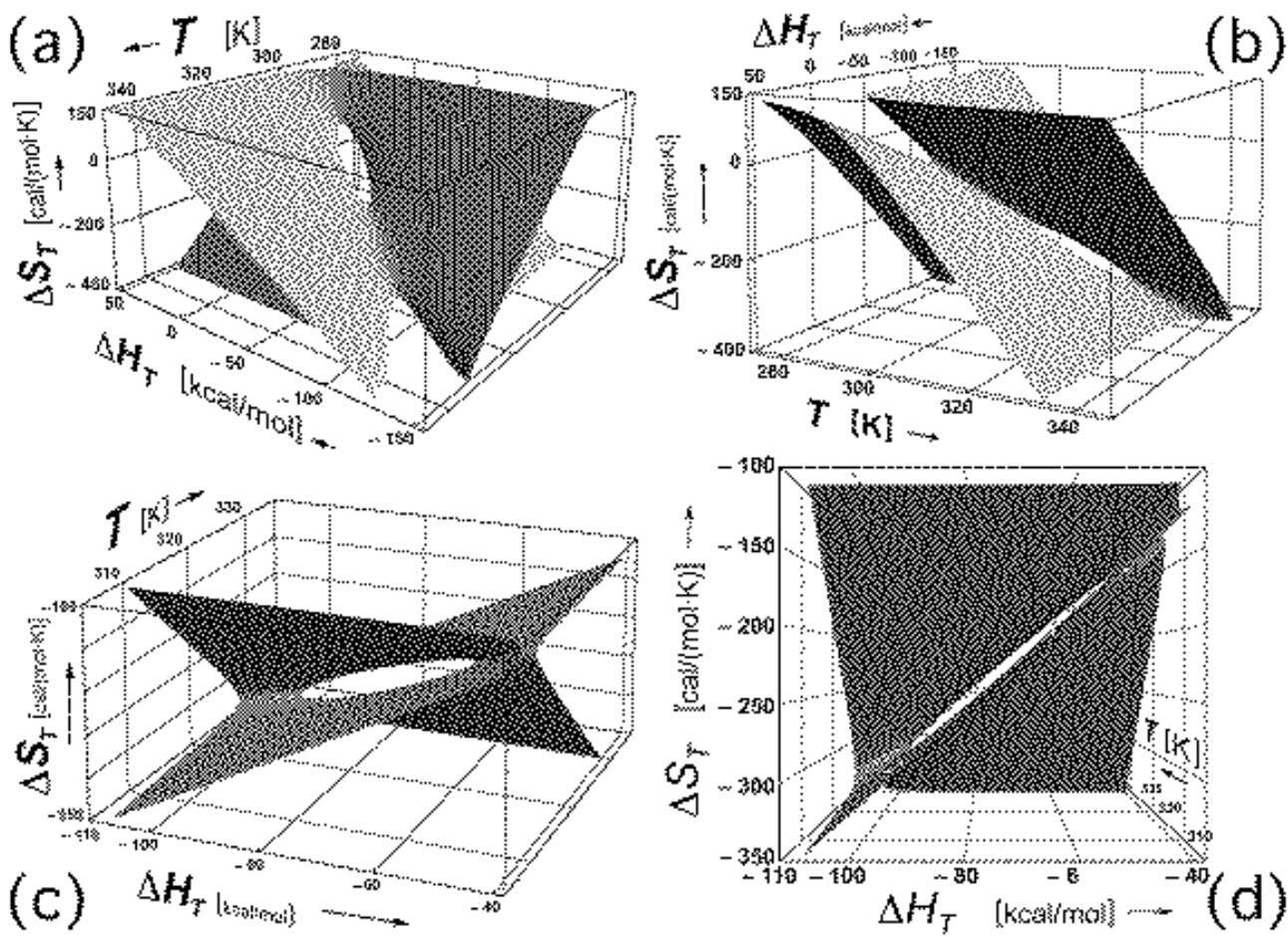

1    **Figure 1.** 3D-Projections $\Delta H_T$ versus $\Delta S_T$ and $T$ of the 5D hyperbolic paraboloid specific of

2    the protein mutant family obtained from ProTherm entry numbers 107-120, pH 7 [3]. (a) and (b):

3    Two relatively narrow and orthogonal wormholes in the central region of the quadric. (c) and (d):

4    The smaller wormhole – also visible in (b) at higher temperatures – hosts the experimental data, *cf.*

5    yellow dots in (d), from which the function was calculated using equations 8 and 9. The narrowness

6    of such wormholes is characteristic for a ubiquitous compensation of $\Delta H_T$ against $\Delta S_T$, as briefly

7    mentioned in the Introduction, and suggests why empirically good, albeit statistically questionable,

8    2D-linear relationships are found in a vast majority of experimental $\Delta H_T$ vs. $\Delta S_T$ correlations.

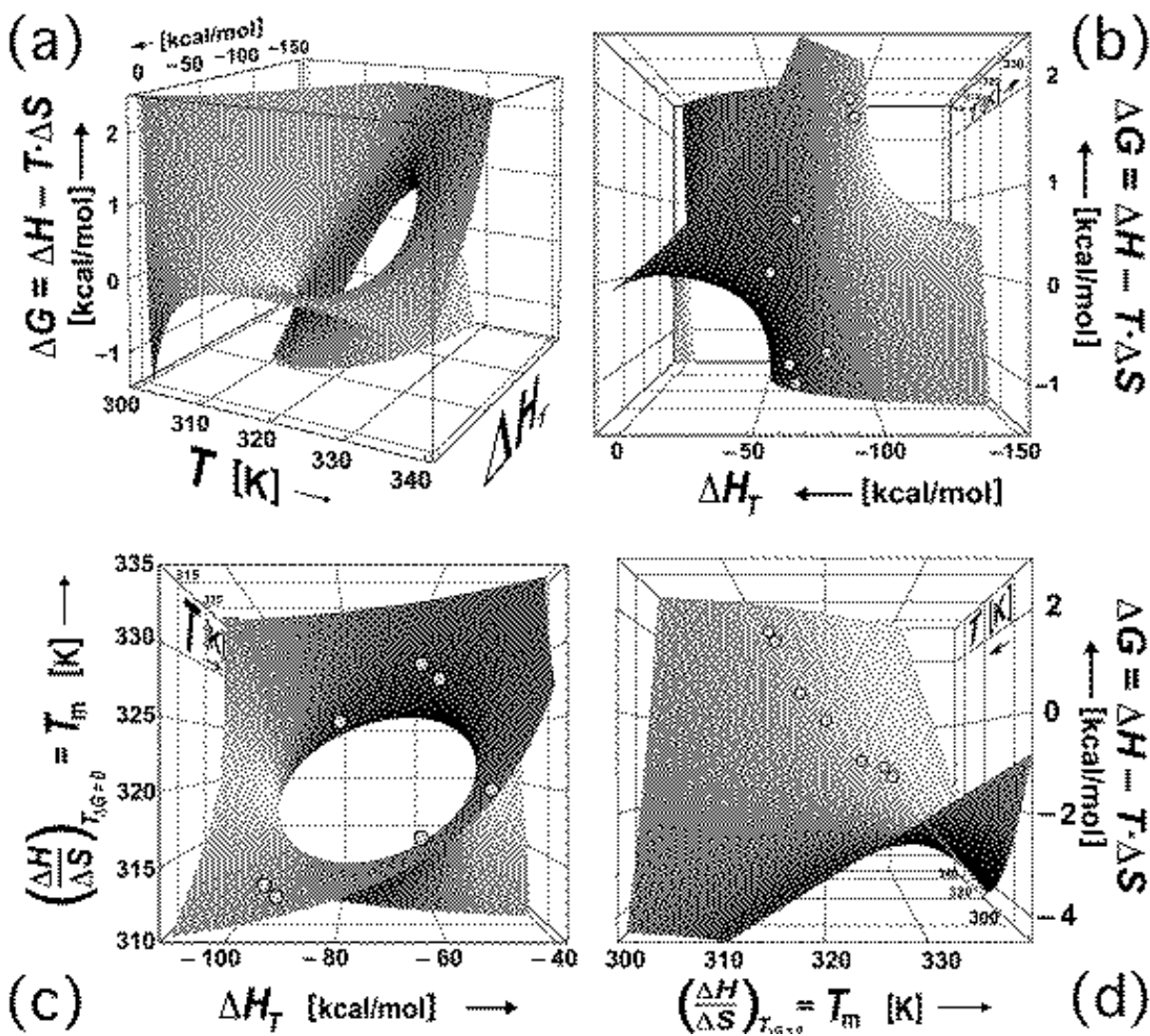

1    **Figure 2.** Other 3D projections of the same group as in Fig. 1. Dimensions (for $\Delta C_p$ = const.):

2    $\Delta H_T$, $\Delta G_T$, $T$ and $T_m$. The yellow dots are the experimental data points $\{\Delta H_{T_m}, \Delta G_{T_{median}}\}$ (b), $\{\Delta H_{T_m},$

3    $T_m\}$ (c) and $\{T_m, \Delta G_{T_{median}}\}$ (d). The yellow line is the linear correlation in $\Delta G_T$ vs. $T_m$ at $T_{median}$ = 320.2

4    K. With the exception of the projection (d), where wormholes are never found and the quality of the

5    linear correlation is best when the data points gather around an average zero free energy, the size of

6    the wormholes in the other (non-$\Delta H_T$ vs. $\Delta S_T$) projections that always host the experimental data

7    points – *cf.* (b) and (c) – precludes any empirical linearity of correlation. All plots generated by

8    MATHEMATICA® (Wolfram Inc.) and edited in PHOTOSHOP® (Adobe).